# Integrated on chip platform with quantum emitters in layered materials

*Sejeong Kim,^,*,1 Ngoc My Hanh Duong,,^,1 Minh Nguyen,1 Tsung-Ju Lu,2 Mehran Kianinia, 1 Noah Mendelson, 1 Alexander Solntsev, 1 Carlo Bradac, 1 Dirk R. Englund, 2 Igor Aharonovich*,1*

[1]School of Mathematical and Physical Sciences, University of Technology Sydney, Ultimo, New South Wales 2007, Australia

[2]Department of Electrical Engineering and Computer Science, Massachusetts Institute of Technology, Cambridge, Massachusetts 02139, USA

^ S. Kim and NMH Duong contributed equally.
E-mail: Sejeong.Kim-1@uts.edu.au ; igor.aharonovich@uts.edu.au

Keywords: 2D materials, single photon emitters, hexagonal boron nitride, integrated photonics

## Abstract

Integrated quantum photonic circuitry is an emerging topic that requires efficient coupling of quantum light sources to waveguides and optical resonators. So far, great effort has been devoted to engineering on-chip systems from three-dimensional crystals such as diamond or gallium arsenide. In this study, we demonstrate room temperature coupling of quantum emitters embedded within a layered hexagonal boron nitride to an on-chip aluminum nitride waveguide. We achieved 1.2% light coupling efficiency of the device and realise transmission of single photons through the waveguide. Our results serve as a foundation for the integration of layered materials with on-chip components and for the realisation of integrated quantum photonic circuitry.

Implementation of quantum networks and photonic processors require interfacing multiple single photons on a chip[1, 2]. For this purpose, efficient integration of quantum light sources and photonic resonators, such as waveguides and cavities, is needed[3, 4]. To this end, photonic integrated quantum circuits have become an attractive research direction, that is poised to deliver compact yet complex solid-state photonic quantum circuitry.

So far, efforts have been predominantly invested in three-dimensional solid-state systems, such as gallium arsenide[5-7] or diamond[8, 9], where the photon source is directly embedded in the material which the resonator is consequently fabricated out of. However, the fabrication of these structures is non-trivial, and alternative hybrid approaches are thus being pursued. One of these hybrid strategies involves transferring the photon source onto pre-made structures fabricated from a different material[10-12]. This has been demonstrated for instance with epitaxially grown quantum dots (QDs). In this case the QD can be pre-characterized and subsequently lifted off and positioned onto a resonator fabricated from conventional semiconductors such as Silicon or Silicon nitride[13-15]. This pick-and-place approach is laborious and requires sophisticated nano-manipulators within scanning electron microscope. It also suffers from limited precision of alignment and positioning of the source with respect to the resonator. Furthermore, single photon emission from the epitaxially-grown QDs can only operate at cryogenic temperatures.

A promising alternative to the above issues is the use of newly emerged single photon emitters in two-dimensional (2D) materials[16, 17]. Due to the two-dimensional nature of the host, they can be transferred onto photonic structures via exfoliation and stamping, reproducibly and in ambient conditions[18]. Indeed, the first work on the coupling 2D materials – namely emitters in layered GaSe[19], to photonic resonators have successfully been realized. However, due to the nature of the emission from these sources, their operation was limited to cryogenic temperatures. A promising alternative is to use the quantum emitters in layered hBN. These sources are ultra bright and operate at room temperature[20-25]. Taking the advantage of these features, in this work we report on the integration of room-temperature hBN quantum emitters with aluminum nitride (AlN) waveguides. We demonstrate transmission of nonclassical light through the waveguide, hence paving the way for future and more complex realisations including photon multiplexing and photonic circuitry on chip.

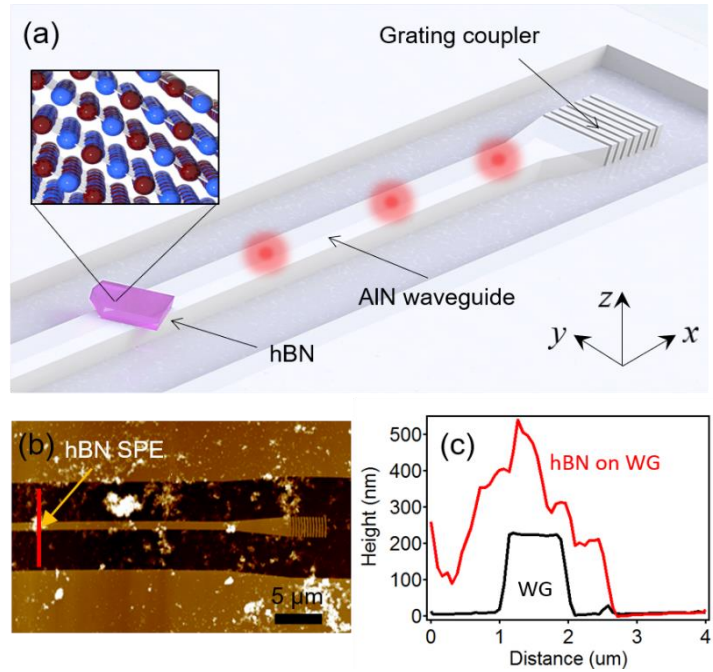

**Figure 1. a)** Schematic view of the hybrid quantum photonic system showing a hBN flake (purple) positioned onto an AlN ridge waveguide (grey). The inset shows the layered van der Waals crystal. **b)** Atomic force microscopy (AFM) image of the hBN-waveguide structure. The position of the hBN emitter onto the waveguide is indicated (yellow arrow). **c)** AFM height measurement of the hBN emitter on the waveguide along the profile following the red line in (b); the ridge measured for the pristine waveguide is also shown (black trace).

Figure 1a depicts schematically the integrated quantum photonic device consisting of the quantum light source and the ridge waveguide terminated with the grating coupler. The inset shows the cross-sectional view of the layered hBN crystalline structure formed by boron and nitride atoms (blue and red spheres respectively). The waveguide is fabricated from aluminum-nitride (AlN) and optimized for single-mode propagation of light with 600-nm wavelength. The width and height of the waveguide are 1 µm and 200 nm, respectively. Nanocrystalline AlN is chosen for the wavaeguide material platform as it has a transparent window from the ultraviolet to the infrared range due to its large bandgap (6.2 eV), while also displaying a low auto-fluorescence, good thermal conductivity and is an electro-optic material[26, 27]. Additionally, the refractive index of AlN ($n = 2.08$) is very close to that of hBN ($n = 2.1$)[28, 29], which desirably minimize light reflection at the AlN/hBN interface. The single photons from the hBN flake placed onto the waveguide (purple solid in Fig. 1a) are coupled and guided through the waveguide itself and detected at its end from the grating coupler. Figure 1b is the atomic force microscopy (AFM) image of the sample showing hBN flakes dispersed on the waveguide. The hBN flake indicated by the yellow arrow is the one characterized in this study. The height profile of the hBN flakes (Figure 1c) indicates partial agglomeration of the hBN flakes over the waveguide resulting in a total height of ~200 nm. .

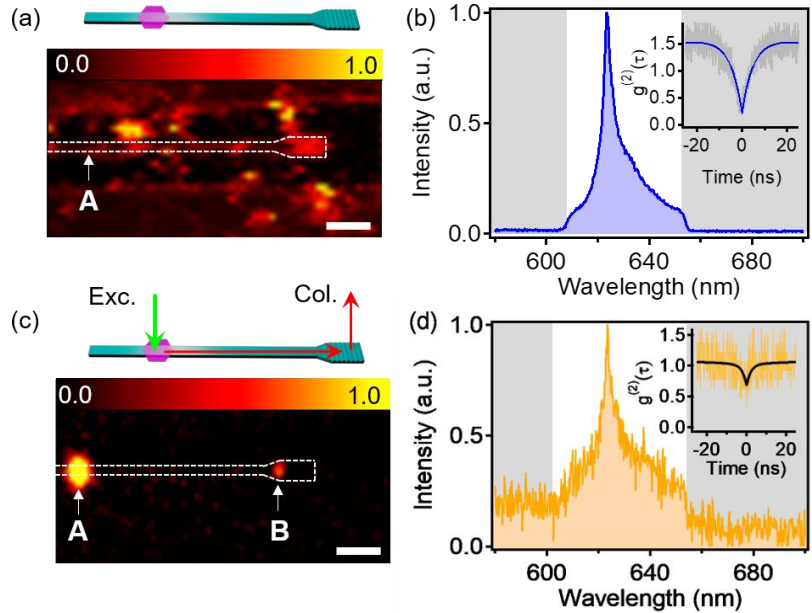

**Figure 2**. Optical characterisation of the hBN-waveguide hybrid structure. **a)** Confocal map of the hybrid system under 532-nm CW laser excitation. **b)** PL spectrum of the hBN emitter and its second-order autocorrelation $g^{(2)}(\tau)$ curve (inset) indicating single-photon emission ($g^{(2)}(0) < 0.5$) . **c)** Confocal map where the 532-nm laser excitation is fixed on the emitter (spot A) and the collection is scanned over the sample; in collection the 532-nm laser is filtered out. The map shows that photons from the emitter couple to the waveguide and are detected at the grating coupler (spot B). **d)** PL spectrum and $g^{(2)}(\tau)$ curves (inset) are collected from the grating coupler (spot B) with excitation fixed at the emitter. Scale bars are 5 µm in both (a) and (c). The collection spectral window is indicated by the unshaded areas in (b), (d). The $g^{(2)}(\tau)$ curves are corrected for background and time jitter.

The confocal photoluminescence (PL) map of the sample is shown in Figure 2a, where the optical excitation was performed with a continuous-wave green laser (wavelength 532 nm) through a 0.9 NA objective. In this standard confocal map, the excitation and collection are from same spot as the sample is scanned using a piezo stage: we refer to this scheme as 'local excitation'. The hBN single-photon emitters (SPEs) on the waveguide were identified and subsequently characterized via this local excitation scheme. Figure 2b shows the zero-phonon line (ZPL) of the hBN SPE at 623 nm with a linewidth of 4.24 nm (full width at half maximum of the fitted Lorentzian). Note, this corresponds to the hBN flake indicated in Figure 1b and 1c. The spectrum was integrated for 60 seconds at an excitation power of 2 mW. The spectrum and second-order autocoreelation $g^{(2)}(\tau)$ measurements in Figure 2b were filtered using a bandpass filter (Semrock, $(630 \pm 28)$ nm) to spectrally reject the Cr luminescence peak from sapphire and to isolate the emitter's ZPL—the regions shaded in grey in Fig. 2b, d display the spectral regions which were filtered out. The insets in Fig. 2b, d show the second-order autocorrelation measurement $g^{(2)}(0)$ performed using a Hanbury-Brown and Twiss interferometer and corrected for background and time jitter. A zero-delay time value of $g^{(2)}(0) = 0.12$, indicates the single-photon nature of the emitter.

Next, the emitter was analysed with a modified collection technique which we refer to as 'non-local collection'—the schematic of which, is shown in Figure 2c. In this scheme, the excitation laser (532 nm) is kept fixed at the location of the hBN SPE (spot A), while collection is acquired over a $30 \times 30$ µm$^2$ area using a scanning mirror; the 532-nm excitation laser is filtered out. Figure 2c shows the resulting confocal scan which reveals luminescence both from the emitter, locally (spot A), and—at lower intensity—from the outcoupling grating (spot B)ized. This demonstrates successful coupling and propagation of single photons from the hBN emitter through the waveguide structure. The bright emission from the spot B in Figure 2c is further analyzed with spectrometer, revealing the same ZPL (623 nm) as from a spot A. Finally, to rule out the potential of an alternative SPE positioned on the waveguide and collected from spot B, we performed additional characterization of this area using the local excitation scheme, and found no other optically-active SPEs on the waveguide, further confirming that the PL from spot B originates from the hBN SPE at spot A.

The intensity of the PL signal measured at the grating coupler (spot B) is 1.2% of the total PL intensity collected from spot A, as shown in figure 2d. Inset is the resulting autocorrelation function, that shows significantly higher $g^{(2)}(0)$ value, most likely due to the increased background compared to the collection via local collection. We attribute this to the system losses which include: (i) coupling efficiency from an emitter to single-mode waveguide, and (ii) extraction efficiency from the grating-coupler. Note, that the propagation loss is negligible as AlN is transparent in the visible region and the propagation length is relatively short (~ 25 μm). Additionally, the waveguide is straight, so no bending losses are considered.

The extraction efficiency of the grating coupler is estimated using finite-difference time-domain (FDTD) simulations. Here, we use the term extraction efficiency to refer to the portion of light that escapes the structure, and is collected through the objective in the optical setup with NA = 1. For the single mode ($\lambda$ = 623 nm) propagating light, the grating coupler used in our experiment has extraction efficiency of 9.8%. This is calculated considering that most of the extracted light is within 20° in the far-field intensity pattern and we used an objective lens with NA = 0.9. The low extraction efficiency is attributed to the high refractive index ($n$ = 1.8) of the sapphire substrate.

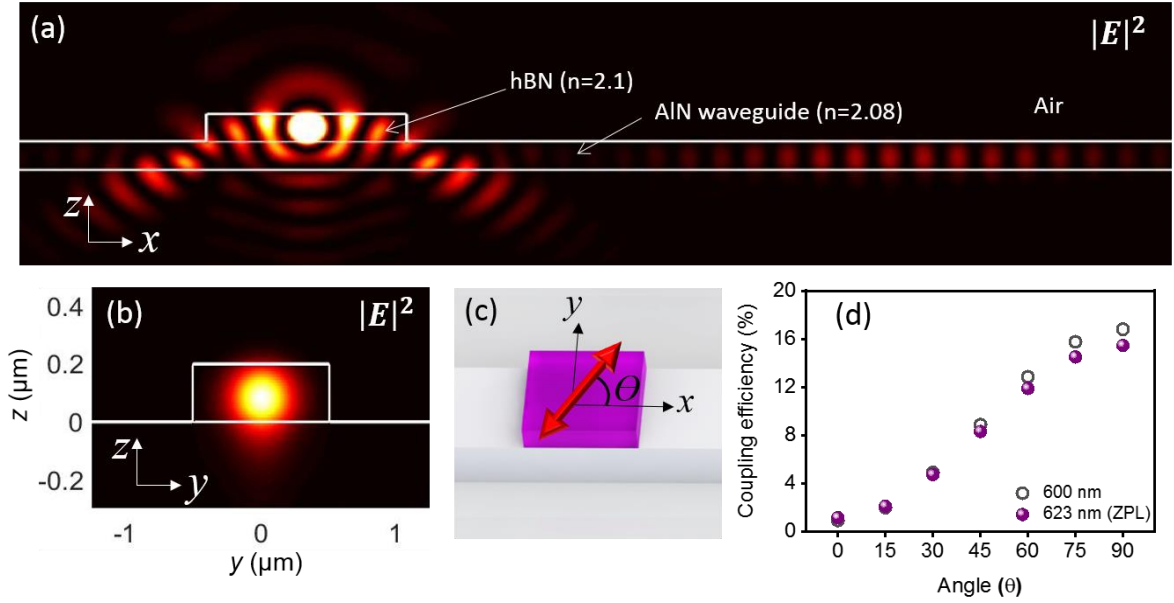

**Figure 3. a)** Three-dimensional FDTD simulation showing the cross section (x-z plane) of the sample, where the emission from the dipole emitter couples to the waveguide. **b)** Mode profile of the waveguide at 623 nm. **c)** Three-dimensional schematic describing the angle of the hBN emitter (red arrow) with respect to the waveguide (the longitudinal axis of the waveguide is along the x-direction). **d)** Coupling efficiency of the light coupling to the waveguide with respect to the emitter's angle.

To complete the analysis on the system efficiency, the coupling efficiency between the emitter and the waveguide is analysed numerically. The hBN flake is simplified using a cuboid with dimensions corresponding to 1000 (width) × 1000 (length) × 200 (height) nm$^3$ and refractive index of 2.1 (in-plane). A dipole emitter with $\lambda$ = 623 nm is located in the center of the flake having polarization along the y-axis orthogonal to the longitudinal axis of the waveguide. Figure 3a shows the cross-sectional view of the electric field intensity showing the light coupled to the AlN waveguide. The coupled light propagates with single gaussian mode as shown in Figure 3b. Additionally, the system is analyzed by varying the emission polarization of the SPE. We focus on the in-plane polarization as, due to the orientation of the flake with respect to the waveguide, this orientation is more relevant to our case. We therefore fix the polarizations in the simulation to the x-y plane while varying the angle $\Theta$ with respect to the waveguide as displayed in Figure 3c. Figure 3d shows the coupling efficiency of the emitter to the waveguide when the wavelength of the dipole emitters are 600 nm and 623 nm, respectively. These two wavelengths were chosen in the simulation because the AlN waveguide was designed and optimized for 600 nm, and the hBN SPE in the experiment has a ZPL at 623 nm. Figure 3d shows that when the emission polarization is along the waveguide direction ($\Theta = 0°$) the coupling is very weak while the efficiency is maximized when the polarization direction is orthogonal to the waveguide ($\Theta = 90°$). The maximum coupling efficiency from the simulation is 15.5% for 623 nm. The hBN SPE in the experiment has a measured emission polarization of $\Theta \sim 90°$; in our estimate of the total coupling efficiency we thus use this value. The total efficiency of the device can be estimated by multiplying all three efficiencies listed above (i, ii, iii), which results in a total value of 1.5%, which matches very well with the value measured experimentally of 1.2%.

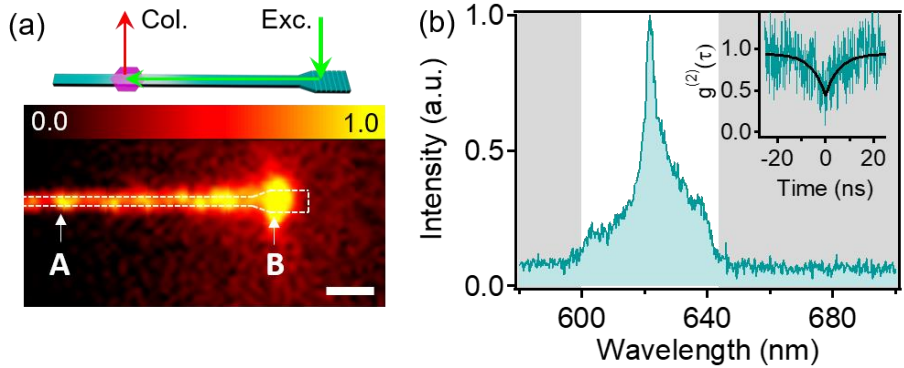

**Figure 4.** Reversed excitation scheme of the system. **a)** Confocal map with excitation fixed on the grating coupler (spot B) and collection acquired from the emitter (spot A). Scale bar is 5 µm. **b)** PL spectrum and $g^{(2)}(\tau)$ function (inset) are collected from the emitter (spot B). The collection spectral window is indicated by the unshaded area in (b). The $g^{(2)}(\tau)$ curves are corrected for background and the time jitter.

Finally, the device was tested in what we refer to as 'reverse collection,', i.e. with the excitation spot fixed at the grating coupler and the collection scanned across the sample area (the laser excitation at 532 nm is filtered out)—the schematic is shown in Figure 4a together with the two-dimensional PL map. The green excitation laser is injected into the waveguide at the grating coupler (spot B) and guided through the waveguide to excite the same SPE that is characterized in Figures 1 and 2. The green laser also excites other photoluminescent defects along the waveguide, yet these reveal broad PL emission which is not related to the investigated SPE. Figure 4b shows the PL spectrum collected from spot A under this reverse collection scheme and displaying the same ZPL wavelength as in Figure 2. The corresponding autocorrelation measurement is shown in the inset of Figure 4b, with $g^{(2)}(0) = 0.42$. The different (higher) $g^{(2)}(0)$ single-photon purity for the reverse scheme in Figure 4b compared to that of the non-local scheme in Figure 2d is attributed to the different collection efficiency for the two different schemes. In Figure 4b, the emission is from a point defect source for which the collection spot size is determined by the NA of the lens. Conversely, in Figure 2d light is scattered by the grating coupler which inevitably has a larger, non-point-like emission. In addition, in the reverse configuration (Fig. 4b) the laser intensity can be increased at will to compensate for losses in the waveguide until we saturate the emitter. However, in the non-local scheme (Fig 2d) while we can still saturate the emitter, photons will be consequently lost while propagating along the waveguide.

To conclude, we demonstrated coupling of room-temperature SPEs in a 2D van der Waals crystal, specifically, hBN to an AlN waveguide. We achieved a coupling efficiency of 1.2% for the hybrid system, which matches the theoretical limit of 1.5% estimated for the structure using FDTD simulations. We also successfully demonstrated, efficient guiding of single photons in a variety of non-local excitation and collection schemes. The hybrid system constitutes the first step towards room-temperature, multi-functional photonic circuitry based on 2D material quantum light sources.

Further optimization can be achieved by improving the coupling efficiency, for instance by engineering the immediate dielectric environment of the emitter through capping with a third material. Better resonator geometries and couplers using the inverse-design principles[30] could also be engineered and fabricated, so that on chip operations can be realized. Finally, we note that research into strain engineered quantum emitters is accelerated[24, 31, 32], which may enable engineering the spectral characteristics of the emitter on demand, and in a precise location on the waveguide.

**Experimental Section**

*Sample preparation:* Hexagonal boron nitride (hBN) nanoparticles in solution (Graphene Supermarket) were transferred onto the waveguides. The whole sample was subsequently annealed at 850 °C for 30 minutes in Argon atmosphere to avoid contamination. After annealing, the system was characterized via confocal microscopy.

*Optical characterization*: The optical characterization of the structures was carried out with a lab-built scanning confocal microscope. A continuous wave (CW) 532-nm laser was used for excitation. The laser was directed and focused onto the sample using a high-numerical-aperture objective lens (NA = 0.9, TU Plan Flour 100X, Nikon). The laser was scanned across the sample using a x-y piezo stage. The collected light was filtered using a 532-nm dichroic mirror (532 nm laser BrightLine, Semrock) and an additional 568-nm long pass filter (LP Filter 568 nm, Semrock). The signal was then coupled into a graded-index fiber, where the fiber aperture served as a pinhole. A fiber splitter was used to direct the light to a spectrometer (Acton SpectraPro, Princeton Instrument Inc.) and to two avalanche photodiodes (SPCM-AQRH-14-FC, Excelitas Technologies) in a Hanbury-Brown and Twiss configuration. In the non-local collection scheme, the excitation was fixed at a specific point and the collection was scanned using a scanning mirror (FSM-300, Newport).

*Numerical calculations:* Photonic simulation was performed by solving Maxwell's equations using the 3D FDTD method (Lumerical software). The size of the simulation domain is 16×5×2 µm$^3$ divided by a 20-nm mesh size. The emitter-waveguide coupling efficiency was simulated with 'Dipole' as a light source while the grating coupler's efficiency was simulated using 'Mode' as a light source.

1. Wang, J.; Paesani, S.; Ding, Y.; Santagati, R.; Skrzypczyk, P.; Salavrakos, A.; Tura, J.; Augusiak, R.; Mančinska, L.; Bacco, D.; Bonneau, D.; Silverstone, J. W.; Gong, Q.; Acín, A.; Rottwitt, K.; Oxenløwe, L. K.; O'Brien, J. L.; Laing, A.; Thompson, M. G., **2018,** *360* (6386), 285-291.
2. Wehner, S.; Elkouss, D.; Hanson, R., *Science* **2018,** *362* (6412).
3. Flamini, F.; Spagnolo, N.; Sciarrino, F., *Reports on Progress in Physics* **2018,** *82* (1), 016001.
4. Bogdanov, S.; Shalaginov, M. Y.; Boltasseva, A.; Shalaev, V. M., *Optical Materials Express* **2017,** *7* (1), 111-132.
5. Kim, H.; Bose, R.; Shen, T. C.; Solomon, G. S.; Waks, E., *Nat. Photonics* **2013,** *7*, 373.
6. Barik, S.; Karasahin, A.; Flower, C.; Cai, T.; Miyake, H.; DeGottardi, W.; Hafezi, M.; Waks, E., **2018,** *359* (6376), 666-668.
7. Singh, A.; Li, Q.; Liu, S.; Yu, Y.; Lu, X.; Schneider, C.; Höfling, S.; Lawall, J.; Verma, V.; Mirin, R.; Nam, S. W.; Liu, J.; Srinivasan, K., *Optica* **2019,** *6* (5), 563-569.
8. Hausmann, B. J. M.; Shields, B.; Quan, Q. M.; Maletinsky, P.; McCutcheon, M.; Choy, J. T.; Babinec, T. M.; Kubanek, A.; Yacoby, A.; Lukin, M. D.; Loncar, M., *Nano Lett.* **2012,** *12* (3), 1578-1582.
9. Evans, R. E.; Bhaskar, M. K.; Sukachev, D. D.; Nguyen, C. T.; Sipahigil, A.; Burek, M. J.; Machielse, B.; Zhang, G. H.; Zibrov, A. S.; Bielejec, E.; Park, H.; Lončar, M.; Lukin, M. D., *Science* **2018,** *362* (6415), 662-665.
10. Türschmann, P.; Rotenberg, N.; Renger, J.; Harder, I.; Lohse, O.; Utikal, T.; Götzinger, S.; Sandoghdar, V., *Nano Lett.* **2017,** *17* (8), 4941-4945.
11. Benson, O., *Nature* **2011,** *480* (7376), 193-199.
12. Khasminskaya, S.; Pyatkov, F.; Słowik, K.; Ferrari, S.; Kahl, O.; Kovalyuk, V.; Rath, P.; Vetter, A.; Hennrich, F.; Kappes, M. M.; Gol'tsman, G.; Korneev, A.; Rockstuhl, C.; Krupke, R.; Pernice, W. H. P., *Nat. Photonics* **2016,** *10*, 727.
13. Davanco, M.; Liu, J.; Sapienza, L.; Zhang, C.-Z.; De Miranda Cardoso, J. V.; Verma, V.; Mirin, R.; Nam, S. W.; Liu, L.; Srinivasan, K., *Nat. Commun.* **2017,** *8* (1), 889.
14. Elshaari, A. W.; Zadeh, I. E.; Fognini, A.; Reimer, M. E.; Dalacu, D.; Poole, P. J.; Zwiller, V.; Jöns, K. D., *Nat. Commun.* **2017,** *8* (1), 379.
15. Kim, J.-H.; Aghaeimeibodi, S.; Richardson, C. J. K.; Leavitt, R. P.; Englund, D.; Waks, E., *Nano Lett.* **2017,** *17* (12), 7394-7400.
16. Aharonovich, I.; Englund, D.; Toth, M., *Nat. Photonics* **2016,** *10* (10), 631-641.
17. Toth, M.; Aharonovich, I., *Annual Review of Physical Chemistry* **2019,** *70* (1), null.
18. Frisenda, R.; Navarro-Moratalla, E.; Gant, P.; Pérez De Lara, D.; Jarillo-Herrero, P.; Gorbachev, R. V.; Castellanos-Gomez, A., *Chemical Society Reviews* **2018,** *47* (1), 53-68.
19. Tonndorf, P.; Del Pozo-Zamudio, O.; Gruhler, N.; Kern, J.; Schmidt, R.; Dmitriev, A. I.; Bakhtinov, A. P.; Tartakovskii, A. I.; Pernice, W.; Michaelis de Vasconcellos, S.; Bratschitsch, R., *Nano Lett.* **2017,** *17* (9), 5446-5451.
20. Tran, T. T.; Bray, K.; Ford, M. J.; Toth, M.; Aharonovich, I., *Nature Nanotech.* **2016,** *11*, 37-41.

21. Mendelson, N.; Xu, Z. Q.; Tran, T. T.; Kianinia, M.; Scott, J.; Bradac, C.; Aharonovich, I.; Toth, M., *ACS Nano* **2019,** *13* (3), 3132-3140.
22. Exarhos, A. L.; Hopper, D. A.; Grote, R. R.; Alkauskas, A.; Bassett, L. C., *ACS Nano* **2017,** *11* (3), 3328-3336.
23. Jungwirth, N. R.; Calderon, B.; Ji, Y.; Spencer, M. G.; Flatt, M. E.; Fuchs, G. D., *Nano Lett.* **2016,** *16* (10), 6052-6057.
24. Proscia, N. V.; Shotan, Z.; Jayakumar, H.; Reddy, P.; Dollar, M.; Alkauskas, A.; Doherty, M. W.; Meriles, C. A.; Menon, V. M., *Optica* **2018,** *5*, 1128.
25. Schell, A. W.; Takashima, H.; Tran, T. T.; Aharonovich, I.; Takeuchi, S., *ACS Photonics* **2017,** *4* (4), 761-767.
26. Liu, X.; Sun, C.; Xiong, B.; Wang, L.; Wang, J.; Han, Y.; Hao, Z.; Li, H.; Luo, Y.; Yan, J.; Wei, T.; Zhang, Y.; Wang, J., *Opt. Express* **2017,** *25* (2), 587-594.
27. Lu, T.-J.; Fanto, M.; Choi, H.; Thomas, P.; Steidle, J.; Mouradian, S.; Kong, W.; Zhu, D.; Moon, H.; Berggren, K.; Kim, J.; Soltani, M.; Preble, S.; Englund, D., *Opt. Express* **2018,** *26* (9), 11147-11160.
28. Kyoungsik Yu, Y. R., Sejeong KIM, and Yeonghoon Jin; Yu, K.; Rah, Y.; Kim, S.; jin, Y., *Opt. Lett.* **2019**.
29. Lee, S.-Y.; Jeong, T.-Y.; Jung, S.; Yee, K.-J., **2019,** *256* (6), 1800417.
30. Molesky, S.; Lin, Z.; Piggott, A. Y.; Jin, W.; Vucković, J.; Rodriguez, A. W., *Nat. Photonics* **2018,** *12* (11), 659-670.
31. Kern, J.; Niehues, I.; Tonndorf, P.; Schmidt, R.; Wigger, D.; Schneider, R.; Stiehm, T.; Michaelis de Vasconcellos, S.; Reiter, D. E.; Kuhn, T.; Bratschitsch, R., *Adv. Mater.* **2016,** *28* (33), 7101-7105.
32. Luo, Y.; Shepard, G. D.; Ardelean, J. V.; Rhodes, D. A.; Kim, B.; Barmak, K.; Hone, J. C.; Strauf, S., *Nature Nanotech.* **2018**.